\documentclass[prl,preprint]{revtex4}   
\usepackage{amsmath,amsfonts}
\usepackage{epsfig}
\usepackage{graphicx}
\newcommand{\dr}{{\rm d}}
\newcommand{\la}{\lambda}
\newcommand{\bea}{\begin{eqnarray}}
\newcommand{\beq}{\begin{equation}}
\newcommand{\eea}{\end{eqnarray}}
\newcommand{\eeq}{\end{equation}}

\begin{document}
\title
{Chirality of wave functions for three coalescing levels} 
\author {WD Heiss} 
\affiliation {Institute of Theoretical Physics, University Stellenbosch\\
National Institute for Theoretical Physics, South Africa }
\begin{abstract}
The coalescence of three levels has particular attractive features.
Even though it may be difficult to realise such event in the laboratory
(three additional real parameters must be adjusted), to take up the 
challenge seems worthwhile. In the same way as the chiral behaviour
of a usual EP can give a direction on a line, the state vectors in
the vicinity of an EP3 provide an orientation in the plane. The
distinction between left and right handedness depends on the
distribution of the widths of the three levels in the vicinity of
the point of coalescence.
\end{abstract}

\maketitle

\section{Introduction}
There is substantial literature during the past decade relating to
exceptional points \cite{kato}, i.e.~points where two eigenvalues 
of an operator coalesce
giving rise to a square root singularity in the spectrum \cite{hesa}.
These singularities are generic and are thus
encountered in virtually all physical problems associated with
eigenvalues. They have been discussed in mechanical problems
\cite{shu}, in optics \cite{berden}, for bound states \cite{hei} and
resonances \cite{mond,korsch} in quantum mechanics and in atomic
physics \cite{lat}. The mutual
influence of neighbouring exceptional points
upon the phase behaviour of the associated wave functions is dealt
with in \cite{sey}. Exceptional points also play a crucial role in
quantum phase transitions \cite{heph}. Chiral behaviour of the
eigenfunction \cite{heha} as well as effects of time reversal
symmetry breaking is discussed in \cite{hahe,mvb}. Experimental
manifestations including chiral behaviour have been achieved with
microwave cavities \cite{darm} and coupled oscillators in electronic 
circuits \cite{st}. A recent more mathematical expos\'e
investigates exceptional points in the context of projective Hilbert
spaces \cite{gunth} while a connection to ${\cal PT}$ symmetric
Hamilton operators \cite{ben} is found in \cite{zno}. In fact,
it is established in \cite{dor} that, for a pseudohermitian 
${\cal PT}$ symmetric Hamiltonian, the onset of spontaneous symmetry breaking by
the wave function happens just at an exceptional point.

In many of the papers quoted above the notation EP was used denoting
the generic exceptional point where two levels coalesce at a square
root branch point. To distinguish it from the non-generic coalescence
of three levels being the subject of the present paper we'll denote
the generic EP by EP2 and we denote by EP3 the specific situation where
three levels coalesce. We note that EPs of higher order have been
implicitly encountered as the coalescence of two or more EP2 in a 
recent investigation of a complex WKB analysis \cite{sorr}.

\section{Three levels coalescing}
The situation of three or more levels coalescing does not seem
to have been investigated in great detail, the reason being that there are too many
parameters needed to enforce such higher order coalescence.
In fact, while two real parameters (one complex parameter) suffice
to invoke the coalescence of two levels, for $N$ levels coalescing
$(N^2+N-2)/2$ real parameters are needed considering complex
symmetric matrices. For $N=3$ it means that three additional real parameters
have to be chosen judiciously to invoke the coalescence of three
levels in the complex plane of some complex parameter. Since, as seen
below, the coalescence of just three levels has particular attractive
features -- for the coordinate systems used conventionally a
distinction between left and right seems possible -- the challenge to 
implement an experimental arrangement may just fall within reach of 
realisation.

We recall that the wave function at a usual EP2
has -- for complex 
symmetric matrices -- a fixed phase relationship of its components
\cite{darm} that can be interpreted as a form of chirality \cite{heha}.
Considering two coupled dissipative oscillators a particular EP2
specifies uniquely which of the two oscillators is leading by
the phase $\pi /2$. In this particular mode, that is at the EP2,
the two oscillators thus specify an orientation in one-dimensional
space by simply placing them on a line and using the convention
that an arrow points from the oscillator with the leading to the one
with the lagging phase \cite{st}.

When three levels are coalescing the structure becomes much richer
in comparison with an EP2, yet much of the structure turns out to be
generalisations that could have been expected in hindsight. Of course, the
simplest form of an operator giving rise to three levels coalescing
is a three-dimensional matrix. Let us assume that any triple of the parameters in
\beq
H_0=\begin{pmatrix} e1 & 0 & 0 \\ 0 & e2 & 0 \\ 0 & 0 & e3 \\
\end{pmatrix}
\quad {\rm and} \quad
H_1=U\begin{pmatrix} o1 & 0 & 0 \\ 0 & o2 & 0 \\ 0 & 0 & o3 \\ 
\end{pmatrix} U^T,
\label{sp}
\eeq
- with $U$ a general three dimensional orthogonal matrix parametrised by
three angles -, is so chosen that the full problem
\beq
H_0+\la H_1
\eeq
has an EP3. If the parameters are all real
(except for $\la $) such EP3 will occur at complex conjugate values of
$\la $. Denoting such point by $\la _c$, the three levels are
connected by a third root branch point (see appendix) 
and there exists the expansion:
\beq
E_j(\la )=E_c+\sum_{k=1}^{\infty }a_k(\root 3 \of {\la -\la _c}\,)^k
\label{exp}
\eeq
where the label $j=1,2,3$ is specified by the first, second or third
Riemann sheet of the third root in the $\la $-plane.
As a consequence, for small values of  $|\la -\la _c|$ the three
complex energies $E_i(\la )$ form an equilateral triangle in the energy plane. The
orientation of the triangle depends on $\arg(\la -\la _c)$ and on the 
parameters of the specific problem (\ref{sp}), which determine the
complex value $a_1$ in (\ref{exp}). Generically
we can order the energies according to their real parts, that is
$\Re (E_1)<\Re (E_2)<\Re (E_3)$ (we dismiss the possibility that two
real parts are equal as non-generic). There are two
basic groups of orientation, depicted schematically in Fig.1: in (a) $\Im(E_2)$ is
smaller than the imaginary parts of the other two energies - in other
words it has the largest width - (recall
that the energies all have negative imaginary parts), whereas in (b)
$\Im(E_2)$ has the largest imaginary part (smallest width).

\begin{figure}[t]
\includegraphics[height=0.22\textheight,clip]{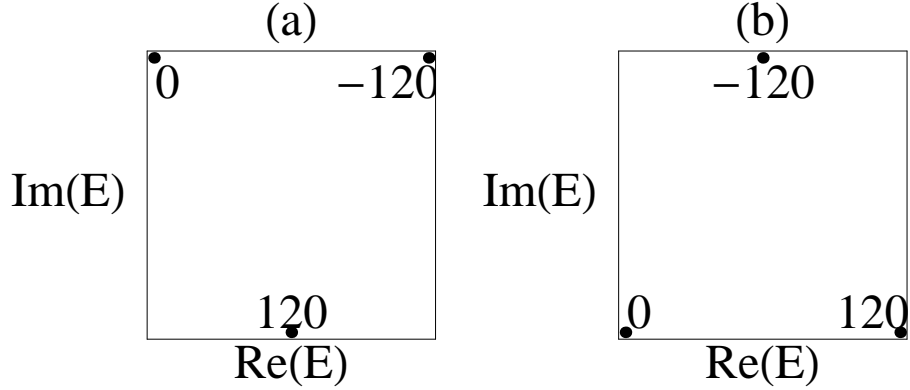}
\caption{The two basic positions of the three levels for small $|\la
  -\la _c|$ in the lower energy
  plane. The drawing is schematic in that the imaginary parts of two
energies are equal: tilting either drawing by less than
$30^0$ still represents the specific case as discussed in the text. 
The lagging (120) and leading (-120) phases of the respective
  wave functions are indicated
  relative to the most left point (smallest $\Re (E)$, denoted $E_1$
  in the main text). The EP3 lies
  in the centre of the equilateral triangles.}
\label{tria}
\end{figure}

Now we turn to the eigenfunctions. It is three eigenfunctions that
become aligned when approaching the EP3. There are the expansions
\beq
|\psi _j(\la )\rangle =|\psi _{EP3}\rangle +
\sum _{k=1} (\root 3 \of {\la -\la _c})^k |\phi _k\rangle .\label{expfun}
\eeq
Note that, as in (\ref{exp}), the label $j=1,2,3$ is again specified by the Riemann sheet
of the third root.
For any $\la \ne \la _c$ the eigenfunctions form the usual
bi-orthogonal complete system, {\it viz.}
\bea
\langle \tilde \psi _i(\la )|\psi _j(\la )\rangle &=&
N_j(\la ) \delta_{i,j} \label{scal} \\
\sum _j\frac{|\psi _j(\la )\rangle \langle \tilde \psi _j(\la )}
{\langle \tilde \psi _j(\la )|\psi _j(\la )\rangle }&=& I.
\label{bas}
\eea
It can be shown (see appendix) that the scalar product (\ref{scal})
vanishes as
\beq
N_j(\la ) \sim \zeta \cdot (\la - \la _c)^{\frac{2}{3}}
\quad {\rm for} \quad \la \to \la _c
\eeq
and similarly
\beq
\langle \tilde \psi _j(\la )|\psi _{EP3}\rangle  
\sim \eta \cdot (\la - \la _c)^{\frac{2}{3}}
\quad {\rm for} \quad \la \to \la _c \label{lim}
\eeq 
 with some constants $\zeta,\eta$ being independent of $j$.

It should be noted that the structure of the eigenvectors at 
an EP3 is slightly more involved as there
are three vectors that coalesce into $|\psi _{EP3}\rangle $ when $\la
\to \la _c$. In view of the result (\ref{lim}) 
the expansion (\ref{expfun}) implies that not only is
$$\langle \tilde \psi _{EP3}|\psi _{EP3}\rangle =0$$
but also $$\langle \tilde \psi _{EP3}|\phi _1\rangle =0$$
where $|\phi _1\rangle $ occurs in the first order term in (\ref{expfun}).

As a consequence, and in contrast to an EP2, the eigenfunction itself
does not {\it a priory} bear a specific chiral phase structure at an
EP3. It is rather in its
immediate neighbourhood where the chiral phase structure is revealed. 
Similar to the reasoning in \cite{heha} an
expansion of $|\psi _{EP}\rangle $ in terms of the normalised basis
\beq
|\chi_j(\la )\rangle =\frac{|\psi _j(\la )\rangle}
{\sqrt{\langle \tilde \psi _j(\la )|\psi _j(\la )\rangle }}
\eeq
yields the proper phase relation. Indeed, while it always
possible for $\la \ne \la _c$ to write
\beq
|\psi _{EP3}\rangle =\sum_{j=1}^3
c_j(\la )\chi _j(\la ) \label{expnorm}
\eeq
identically in $\la $, it is in particular for $\la \sim \la _c$
\beq
\begin{pmatrix}
c_1(\la )  \\ c_2(\la ) \\ c_3(\la )\end{pmatrix}\sim 
\xi \root 3 \of {|\la -\la_c|} \begin{pmatrix}
e^{2 i \pi j_1/3} \\ e^{2 i \pi j_2/3} \\ e^{2 i \pi j_3/3}
\end{pmatrix} 
\label{phas}
\eeq 
with $j_1,j_2,j_3$ being a fixed permutation of $0,1,2$ specified 
below and $\xi $ a complex constant.
Recall that the normalised state vectors $\chi_j(\la )$
blow up as $1/\root 3 \of {\la -\la _c}$ thus yielding
a finite expression for $|\psi _{EP3}\rangle $ in (\ref{expnorm}). The result is
due to the analytic structure of the third root branch point and is
thus independent of a particular basis. As for an EP2, the
fixed phase relations between the eigenstates in the immediate vicinity
of the EP3 are of significance. This result is of course related to
(\ref{expfun}), the appealing feature of (\ref{phas}) lies in its
similarity with the situation for an EP2. Note, however, that for an
EP2 the phase difference is $\pi/2$ while here it is $2\pi/3$. This
means that, in contrast to the EP2, where only a fourfold loop
around the singularity restores the eigenfunctions, for the EP3
the eigenvectors are retrieved
after three loops in the $\la $-plane just like for 
the corresponding energies.

The interesting question is now the precise association of the phases
of the eigenfunctions with the complex values of the energies, that is
with the frequencies and their widths. This association is illustrated
in Fig.1. Choosing as reference point the eigenvector of the energy
with smallest real part (frequency), then the eigenstate of the next
frequency has a lagging phase of $120^0$ if (and only if) the width is
larger than the other two; the phase of the eigenstate with largest
frequency would then be leading by $120^0$. In turn, if the width of
the middle frequency is smallest, the role of leading and lagging phase
is swapped among the two states with the larger frequencies. This
result is demonstrated below in a specific setting and confirmed
numerically in numerous general examples.

The strict phase relations associated with the positions of the
frequencies and widths of the three levels allows the interpretation
of clear distinction between a left hand and a right hand
helix. Fig.2 illustrates how a particular right handed helix is generated:
in a three dimensional coordinate system the points
$(\cos(\Phi_k),\sin(\Phi_k),\Re(E_k))$ with 
$\Phi _{1,2,3}=(0^0,120^0,240^0)$ invoke an oriented helix.
The example refers to the case where the width of the middle frequency
is largest, it generates a right handed helix. In turn, if the width
of the middle frequency is smallest, the helix will be left handed.

\begin{figure}[t]
\includegraphics[height=0.22\textheight,clip]{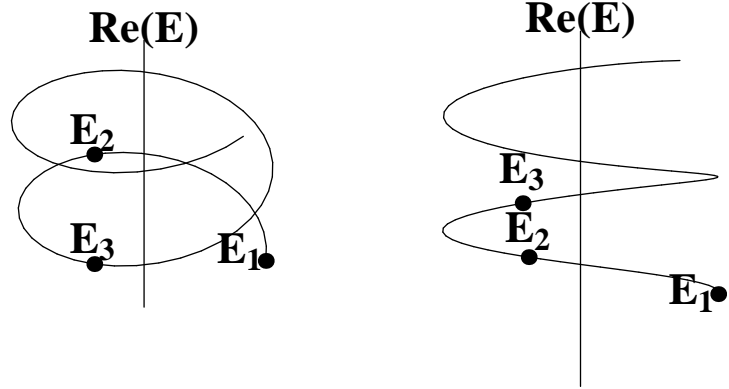}
\caption{Two perspective views of the 
three dimensional helix when the width of $E_2$ is larger
than the other two. The coordinates of the points denoted by $E_k$
are $(\cos(\Phi_k),\sin(\Phi_k),\Re(E_k))$ with 
$\Phi _{1,2,3}=(0^0,120^0,240^0)$.}
\label{hel}
\end{figure}

\section{Special setting}

The all important question is: can a threefold coalescence be arranged
in the laboratory and are the energies and phases amenable to
measurement? One suggestion could be a setting
similar in spirit to the microwave experiment for an EP2. I could
imagine a setup as illustrated in Fig.3. Again, this can be simulated
by the simple matrix
\beq
H=\begin{pmatrix} e_1 & s1 & s3 \\ s1 & e_2 & s2 \\ s3 & s2 & e_3
\end{pmatrix}
\eeq
where the $s_j$ give the couplings (gaps in Fig.3) and the $e_j$ are
the (complex) energies. To facilitate matters, we consider in particular
the problem
\beq
H_0+s1\,H_1=
\begin{pmatrix} e_1 & 0 & s3 \\ 0 & e_2 & s2 \\ s3 & s2 & e_3
\end{pmatrix}+
\la \begin{pmatrix} 0 & 1 & 0 \\ 1 & 0 & 0 \\ 0 & 0 & 0
\end{pmatrix}
\label{hampart}
\eeq
and choose the couplings $s_2$ and $s_3$ such that $H_0$ has a
threefold coalescence. The choice

\begin{figure}[t]
\includegraphics[height=0.16\textheight,clip]{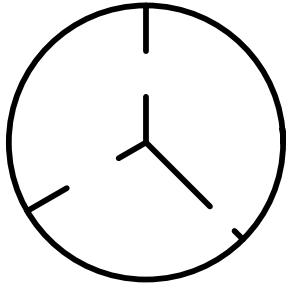}
\caption{Suggested arrangement, arbitrarily chosen, of a microwave cavity to
invoke a threefold coalescence of levels. As for an EP2 the opening of
the gaps and some Teflon piece in one or two of the chambers can be
used as parameters to steer the system into an EP3. 
To avoid systematic degeneracies of the levels there should be no symmetry
in the geometry of the cavity.}
\label{circ}
\end{figure}

\bea
s2&=&\pm \sqrt{-\frac{(e_1-2e_2+e_3)^3}{27(e_1-e_2)}}  \nonumber \\
s3&=&\pm \sqrt{+\frac{(-2e_1+e_2+e_3)^3}{27(e_1-e_2)}}
\label{partic}
\eea
achieves this goal with $$E_c^{(1,2,3)}=\frac{1}{3}(e_1+e_2+e_3). $$
The perturbation by $\la $ splits the coalescence into three levels and
we obtain to lowest order (with $\la _c=0$)
\bea
E_j=E_c&+& 
\frac{2^{1/3} \sqrt{-(-2e_1+e_2+e_3)(e_1-2e_2+e_3)}}{3(e_1-e_2)^{1/3}}
\root 3 \of {|\la-\la _c|}
\exp (2i(j-1)\pi/3)  \nonumber \\
&+&O((\la-\la _c)^{4/3}) 
\eea
and, using the notation (\ref{expfun}), the corresponding
(unnormalised) eigenvectors are
\bea
|EP\rangle &=&
\begin{pmatrix} \frac{\sqrt{-2e_1+e_2+e_3}}{\sqrt{3(e_1-e_2)}} \\ 
i  \frac{\sqrt{e_1-2e_2+e_3}}{\sqrt{3(e_1-e_2)}}  \\ 1 \end{pmatrix} \label{epf}
\\
|\phi _1^j\rangle &=&
\begin{pmatrix} i 2^{1/3} \frac{\sqrt{e_1-2e_2+e_3}}{\sqrt{3}(e_1-e_2)^{5/6}}  \\ 
 - 2^{1/3} \frac{\sqrt{-2e_1+e_2+e_3}}{\sqrt{3}(e_1-e_2)^{5/6}}   \\ 0
\end{pmatrix} \exp (2i(j-1)\pi/3), \quad j=1,2,3
\eea
and similar algebraic expressions for the higher orders. 

The essential
point here is the fact that - to lowest order - 
the three complex eigenvectors $|\phi _1^j\rangle $ differ
only by a well defined phase; the precise specification of the particular
eigenvector including its all important phase is given by its association 
with the levels $E_j$ as indicated in Fig.1: the width of the second
energy of the ordered frequencies characterises the orientation of the
phases $\exp (2 i j\pi/3)$; if $\Im E_2$ is smallest or largest a left 
or right hand helix is obtained, respectively.

\section{Conclusion}
A measurement of all three frequencies and their associated widths will thus
predict the phase of the wave function in the immediate vicinity of
the EP3. We recall that the single eigenstate $|\psi _{EP3}\rangle $
does not bear such information. The relevant phases sit in the part
of the eigenstate that is switched on when moving away from the EP3
in the $\la $-plane. An experimental verification of this subtle behaviour
would thus lead to a clear chiral characterisation in three
dimensional space. We stress that there is no {\it a priory}
handedness in a setting as suggested for instance by Fig.3, neither is
there any involvement of weak interaction. There is, however, the
direction of time that specifies the various widths of a dissipative
system. In other words, we here suggest that the arrow of time can invoke
chirality.

\section{Acknowledgement}

The author gratefully acknowledges constructive comments from all colleagues
working at microwave cavity experiments at the TU Darmstadt.
\bigskip 

\section{Appendix}
For an $N$-fold coalescence of eiegnvalues the $N$ levels are
connected by a branch point of $N$-th order. This follows from the fact
that - for an $N$-dimensional matrix of the form $H_{\la }=H_0+\la H_1$ - the
determinant of $|H_{\la }-I E|$ vanishes linearly in the variable $\la
$ while the requirement of $N$ coalescing levels entails an $N$-fold
vanishing in the variable $E$, that is the following set of equations
is to be satisfied simultaneously
$$ \frac {\dr ^k}{\dr E^k}|H_{\la }-I E|=0, \quad k=0,\ldots,N-1. $$
This is possible only if
$$E(\la )=E_c+\sum _{m=1}^{\infty} c_m(\root N \of {\la -\la _c}\,)^m.$$
Here the $N$ eigenvalues are given by the values upon the $N$
sheets of the $N$-th root.

The $N$ eigenstates coalesce likewise into one eigenvector. There is
the expansion
$$|\psi (\la )\rangle=|\psi _{EP\,N}\rangle+
\sum _{m=1}^{\infty} |\phi _m\rangle (\root N \of {\la -\la
  _c}\,)^m.$$
Note, however, that for the scalar product the following behaviour prevails
$$  \langle \psi (\la )|\psi _{EP\,N}\rangle \sim (\la -\la
_c)^{\frac{N-1}{N}} $$
as follows from considering
\bea (E(\la ))-E_c)\langle \psi (\la )|\psi _{EP\,N}\rangle &=&
\langle \psi (\la )|H_{\la }-H_{\la _c}|\psi _{EP\,N}\rangle \nonumber
\\
&=& (\la -\la _c)\langle \psi (\la )|H_1|\psi _{EP\,N}\rangle .\nonumber
\eea
The right hand side vanishes linearly when $\la \to \la _c$ while
$(E(\la ))-E_c)\sim \root N \of {\la -\la _c}$. Note that these
analytic properties imply the relations
$$\langle \psi _{EP\,N}|\phi _m\rangle =0 \quad {\rm for}\quad
m=1,\ldots,N-1 $$ and
$$\langle \phi _m|\phi _{m'}\rangle =0 \quad {\rm for}\quad
m+m'\le N-2 .$$

\end{document}